\documentclass[12pt]{iopart}

\usepackage{aas_macros}
\usepackage{epsf}
\usepackage[dvips]{graphics,color}

\def\ion#1#2{#1\,{\sc #2}}
\newcommand{\lam}{$\lambda$}

\begin{document}

\title[The
  CHIANTI Atomic Database]{The
  CHIANTI Atomic Database}

\author{P R Young$^{1,2}$, K P Dere$^1$, E Landi$^3$, G Del Zanna$^4$
  and H E Mason$^4$}

\address{$^1$ College of Science, George Mason University, Fairfax, VA
  22030, USA}
\address{$^2$ NASA Goddard Space Flight Center, Code 671, Greenbelt,
  MD 20771, USA}
\address{$^3$ Department of Climate and Space Sciences and
  Engineering, University of Michigan, Ann Arbor, MI 48109, USA}
\address{$^4$ DAMTP, Centre for Mathematical Sciences, University of
  Cambridge, Wilberforce Road, Cambridge, CB3 0WA, UK}

\begin{abstract}
  The CHIANTI atomic database was first released in 1996 and has had a
  huge impact on the analysis and modeling of emissions from astrophysical
  plasmas. The database has continued to be updated, with version~8
  released in 2015. Atomic data for modeling the emissivities of 246
  ions and neutrals are contained in CHIANTI, together with data for
  deriving the ionization fractions of all elements up to zinc. The different types
  of atomic data are summarized here and their formats
  discussed. Statistics on the impact of CHIANTI to the astrophysical
  community are given and
  examples of the diverse range of applications are presented.
\end{abstract}

%
%
%
%
%

\section{Introduction}

The CHIANTI atomic database was first released in 1996 \cite{dere97}
in time for the launch of the Solar and Heliospheric Observatory
(SOHO; Domingo et al.~\cite{soho}).
This mission contained three ultraviolet
spectrometers, and there was a community need for a freely-available
atomic database and software package that would enable researchers to
apply the latest atomic data-sets to analyze these data. The solar
vacuum ultraviolet spectrum (roughly 100--2000~\AA) is rich in
emission lines formed over the temperature range $10^4$--$10^7$~K, and
much activity took place from the 1960's through to the 1980's in
first identifying these lines and then calculating atomic data that
could be used to interpret the lines. By the early 1990's atomic data
were available for most of the abundant ions, yet the data and
software to compute emissivities were not
easily accessible to most researchers. K.~Dere (Naval Research
Laboratory, USA), H.~Mason (Cambridge University, UK) and B.~Monsignori-Fossi
(Arcetri Observatory, Italy) devised an atomic database that would be
freely-available to the community with software written in the
Interactive Data Language (IDL) that was, and still is, widely used in
Solar Physics. Since 1996 the CHIANTI database has continued to
expand, including detailed  coverage of the X-ray wavelength
region (1--100~\AA), atomic data for computing the ionization
balance of electron-ionized plasmas, and a new \emph{Python}-based
software package. Version~8 of CHIANTI was released
in September 2015 \cite{chianti8} and features many new data-sets that
have resulted from recent large-scale atomic calculations, with
particular focus on the coronal iron ions.

The present article describes the database and demonstrates the scope
and impact of the project. A summary of each of the data-sets
currently in CHIANTI is given, and examples of applications are discussed.

\section{Scope and impact}

\begin{figure}[h]
\centerline{\epsfxsize=6in\epsfbox{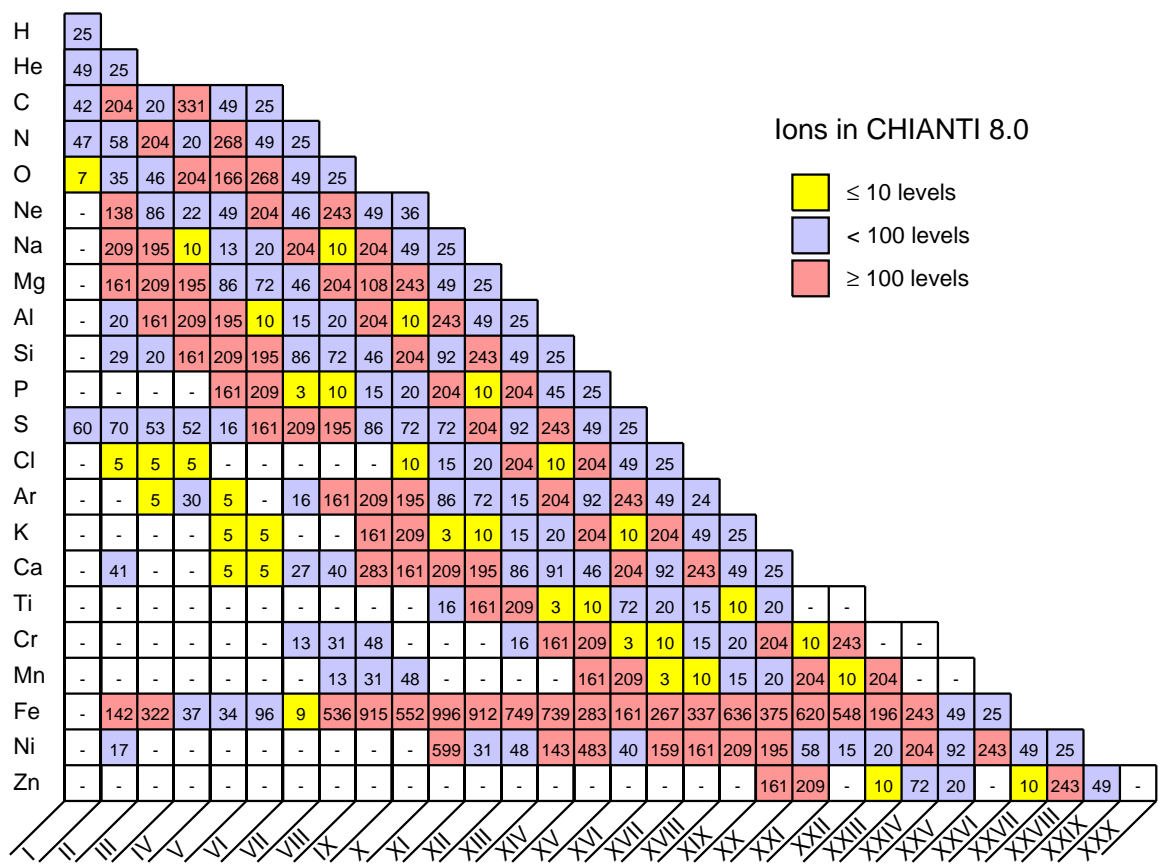}}
\caption{A table showing the elements and ions in CHIANTI 8. The
  number in each box indicates the number of atomic levels in the CHIANTI
  model for that ion. CHIANTI  contains elements up to zinc, and
  low abundance elements such as Li, Be and B are not included.}
\label{fig.table}
\end{figure}

CHIANTI contains atomic data for positively-charged ions and
neutrals and 
Figure~\ref{fig.table} shows the species currently found in
the  database.  
Successive versions of CHIANTI have both
expanded the coverage of the database and increased the sizes of the
atomic models. As illustrated by the pink boxes in
Figure~\ref{fig.table}  a significant amount of effort has been
expended in providing large atomic models for the coronal iron ions
(\ion{Fe}{viii--xxiv}) that are critical for many studies in
astrophysics. 

\begin{figure}[h]
\centerline{\epsfxsize=15cm\epsfbox{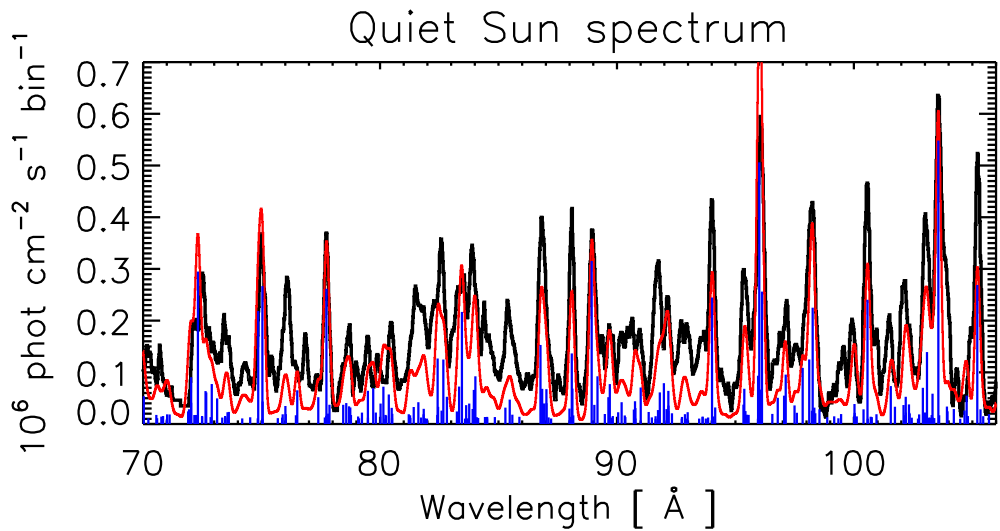}}
\caption{The black line shows an observed quiet Sun spectrum
  \cite{manson72}, and the red line shows the synthetic CHIANTI
  spectrum. Blue lines show the locations of the strongest transitions.}
\label{fig.spec}
\end{figure}

The need for detailed atomic models for the iron ions is illustrated
in Figure~\ref{fig.spec}, which shows an observed quiet Sun spectrum
from Manson~\cite{manson72} in the wavelength region
70--106~\AA. Following the addition of new atomic data for the $n=4$
levels of the coronal iron ions and new line identifications by Del
Zanna~\cite{delzanna:12} in CHIANTI 8 \cite{chianti8} there is
now much improved agreement with the observed spectrum. This
wavelength region is now mostly complete in terms of atomic
transitions, however many of  the emission lines only have theoretical
wavelengths.

\subsection{Data assessment}

The CHIANTI team provide a single atomic model for each species in the
database, with the best atomic data selected from the published
literature. The assessment process is therefore a crucial component of
the team's work, and we identify three key elements:
\begin{enumerate}
  \item graphical assessment of electron excitation collision
    strengths;
  \item comparisons of new data-sets with previous data-sets; and
  \item benchmark comparisons of emission line emissivities against
      observed spectra.
\end{enumerate}
The electron collision strengths are the most important data-set in
CHIANTI and a graphical procedure is applied to each atomic transition
in order to
identify anomalies such as typographic errors, discontinuities and,
most critically, that the data points tend to the high temperature
limit (see Sect.~\ref{sect.exc} and Dere et al.~\cite{dere97}). Where
multiple data-sets are available, or a new data-set is being
considered, then parameters are compared with a particular focus on
the most important transitions for the ion.

The CHIANTI team have performed many benchmark studies to assess the
accuracy of the atomic models, which serve both to validate the models
and also to identify areas where improved data are required. The assessments include detailed analyses of
specific solar spectra, such as those performed by Young et
al.~\cite{1998A&A...329..291Y}, Landi et al.~\cite{landi02}, Landi \& Phillips~\cite{landi06}, Landi \& Young~\cite{landi09}, Young \&
Landi~\cite{young09-cool} and Del Zanna~\cite{2012A&A...537A..38D}. In
addition benchmark studies for the coronal 
iron ions have been performed by Del Zanna~\cite{dz-fe7}, Del
Zanna~\cite{2009A&A...508..513D}, Del Zanna et al.~\cite{2004A&A...422..731D},
Del Zanna~\cite{2010A&A...514A..41D}, 
Del Zanna \& Mason~\cite{2005A&A...433..731D}, and Del
Zanna~\cite{2011A&A...533A..12D} for ions \ion{Fe}{vii--xiii},
respectively. 

\subsection{Impact of CHIANTI}

\begin{figure}[h]
\centerline{\epsfxsize=12cm\epsfbox{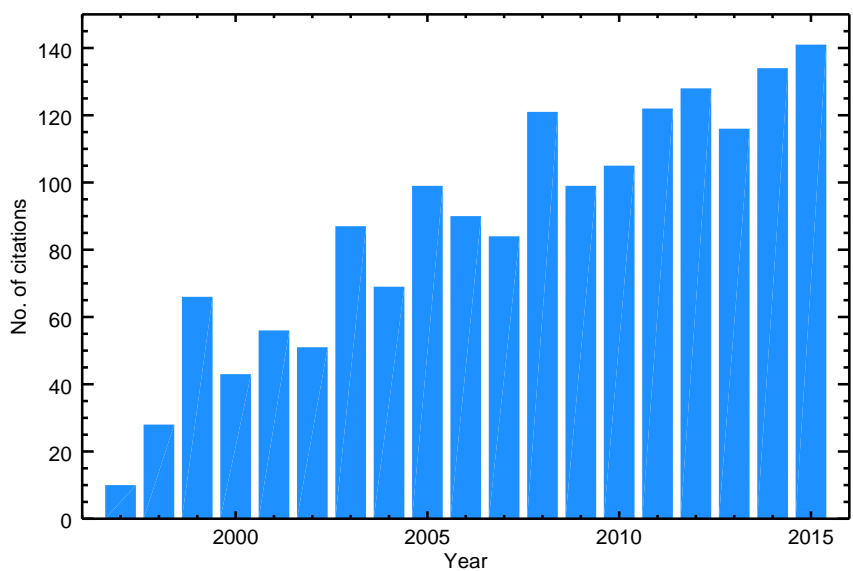}}
\caption{The number of citations the CHIANTI papers have received as a
  function of time. The value for 2015 is an extrapolated value based
  on citations up to September.}
\label{fig.cit}
\end{figure}

\begin{figure}[h]
\centerline{\epsfxsize=12cm\epsfbox{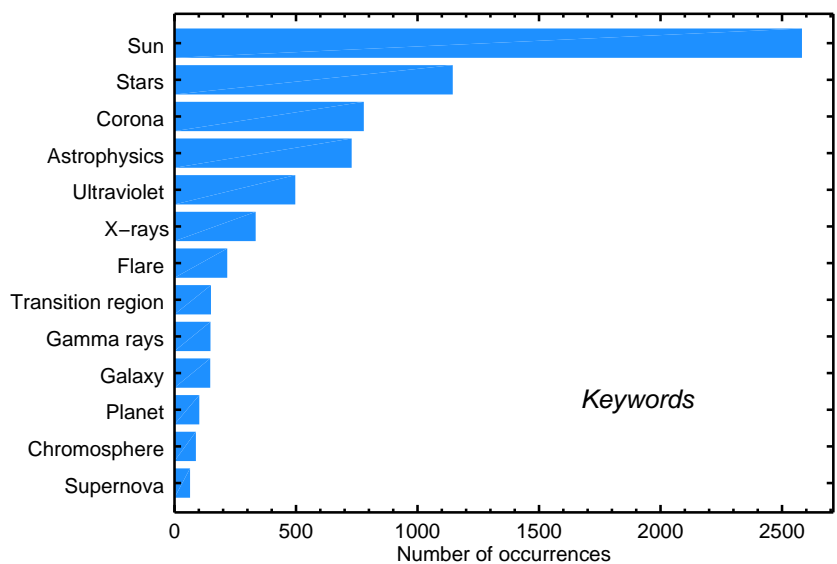}}
\caption{A histogram showing the most commonly-used keywords for
  papers citing the CHIANTI papers.}
\label{fig.hist}
\end{figure}

One measure of the success of CHIANTI is the number of
citations the CHIANTI papers have received. There are 14 papers in
all, nine of which describe the database and its updates, and the
remaining papers present comparisons of CHIANTI with observed
spectra. As of 2015 September 11, the CHIANTI papers had received 2091
citations from 1606 unique
papers\footnote{http://www.chiantidatabase.org/chianti\_ADS.html}, and
the number of citations with time is shown in Figure~\ref{fig.cit}. An
indication of the range of applications of CHIANTI can be made by
identifying the most commonly-used keywords used by papers citing the
CHIANTI papers, and these are shown in Figure~\ref{fig.hist}. ``Sun''
is the most common, reflecting CHIANTI's origins in the Solar Physics
community, but ``Stars'' and ``Astrophysics'' are also prominent. We
also highlight the large number of occurrences for both ``X-rays''
and ``Gamma rays'' reflecting the importance of CHIANTI for these
areas even though CHIANTI was originally developed for modeling
ultraviolet spectra. 

CHIANTI was originally designed for the needs of spectroscopists
and it is now a widely-used tool for both solar and astrophysical
spectroscopists. However CHIANTI is also used by many researchers in
the plasma modeling community who need to be able to predict the
radiative 
emissions from the plasma they model. Examples include the
Hydrodynamic Radiation code (HYDRAD; \cite{bradshaw03a,bradshaw04})
for modeling solar coronal loops, the
Corona-Heliosphere code (CORHEL; \cite{lionello09}) and Alfven-Wave
Solar Model (AWSoM; \cite{holst14}) 
for modeling the
global solar coronal emission,  and the Radiative Transfer and
Hydrodynamics code (RADYN; \cite{allred15}) code for modeling solar and stellar 
flares. Astrophysics modeling codes that use CHIANTI include
TARDIS for modeling emission from supernovae \cite{kerzendorf14}, the
Monte Carlo Simulations of Ionised Nebulae (MOCASSIN) photoionization
code \cite{ercolano08}, and the code of
Richings et al.~\cite{richings14} for modeling the chemistry of the
interstellar medium in galaxy simulations.
The quality of the atomic data in CHIANTI has been recognised by other
plasma codes that often choose to directly ingest the CHIANTI atomic
data. Examples, include \emph{Cloudy} \cite{Lykins15}, XSTAR
\cite{bautista01} and ATOMDB \cite{smith01}.

Another use of CHIANTI is by space instrument teams who use it for deriving
instrument response functions or for calibrating their instruments. An
example is the Atmospheric Imaging Assembly (AIA; \cite{aia}) on board the Solar
Dynamics Observatory (SDO; \cite{sdo}) which obtains high resolution images of the Sun
at ultraviolet and extreme ultraviolet wavelengths through narrow-band
imaging. The response of the instrument's filters to the plasma
temperature depends on the spectral content of the bandpasses and for
these CHIANTI is used \cite{boerner14}. The radiometric calibration of
spectrometers can be checked by using the spectra themselves, for
example certain ratios are known to be insensitive to plasma
conditions. Revisions to the calibration of the EUV Imaging
Spectrometer (EIS; \cite{culhane07}) on board the \emph{Hinode} spacecraft
\cite{kosugi07}  were made by Del Zanna~\cite{dz-cal} and Warren et al.~\cite{warren14} by
making use of the CHIANTI database.

\section{Level balance within an ion}

Ions in low-density astrophysical plasmas are mostly in their ground
electronic state and they are occasionally excited through collisions
with free electrons.
The
excited states
decay almost immediately back towards the ground through spontaneous
radiative decay with the emission of photons. The situation is
complicated by metastable levels and cascading. Metastable levels are
mostly found  within the ion's ground
configuration and they gain significant population with increasing
density, providing additional excitation routes. Cascading is the
process by which a level decays towards the ground level not in a
single jump, but via other excited states. As two distinct atomic processes
are at work in the excitation-decay process then the  system is not in thermodynamic
equilibrium and so the level populations of the ion have to be modeled
by detailed balance of all the levels together, with atomic rates
required for all levels in the model.

An important simplification is that ionization and recombination
processes are generally much less frequent that the excitation and
decay processes, and so the ionization balance equations are treated
separately from the level balance equations. So, although we need the
quantity $N_j$, the number density of ions in a certain atomic state
$j$, to compute an emission line's intensity, we separately compute the
quantity $n_j=N_j/\left(\sum_j N_j \right)$ from level balance
equations that consider only processes affecting levels within the
ion. The full quantity $N_j$ is computed later by including the
results of the separate ionization balance equations.

The level balance equations are considerably more complex than the
ionization balance equations, and  the principal purpose of CHIANTI
is to provide the atomic data for 
accurately solving these level balance equations. Electron excitation
rates and spontaneous radiative decay rates are the principal
data-sets, but there are additional processes that need to be included
for many ions and these are described in later sections.




Collisionless processes are described by rates (units: s$^{-1}$)
while collision processes are described by rate coefficients (units:
cm$^3$~s$^{-1}$) so, for example, the total number of transitions leaving a
state $j$ to enter a lower state $i$ through radiative decays and electron de-excitations is
$N_jA_{ji} + N_j N_{\rm e} C_{ji}$ where $N_j$ is the number density of
particles in the atomic state $j$, $A_{ji}$ is the radiative decay
rate, $N_{\rm e}$ is the electron number density and $C_{ji}$ is the
electron rate coefficient. The level balance equation for the level
$j$,  assuming only radiative decay and electron excitation are the
important processes, is:
\begin{eqnarray}
 \sum_{k>j} N_kA_{kj} + N_{\rm e}\sum_{i\ne j} N_iC_{ij}  & = &
 \sum_{i<j} N_jA_{ji} + \sum_{k\ne j} N_j N_{\rm e} C_{jk} \\
 {\rm [population~into~level}~j & = & {\rm population~out~of~level}~j] 
\end{eqnarray}
In practice, for any specific level many of the terms in this equation
will be zero or negligible. For example, an excited, non-metastable
level will have radiative decay rates orders of magnitude larger than
upwards or downwards electron excitation rates and so the latter will
be negligible. For an excited metastable level in the ground configuration,
however, many of the rates will be comparable in size and so all
processes have to be included.

Solving the level balance equations yields the set of
$n_j$ values for a specified electron number density and temperature
($T$), and the \emph{emissivity} of an 
emission line resulting from a $j\rightarrow i$ decay occuring within
an ion X$^{+q}$ of element X is written as
\begin{equation}
  \epsilon_{ij}(N_{\rm e},T) = {N({\rm X}) \over N_{\rm H}}
             {N_{\rm H} \over N_{\rm e}}
  {N({\rm X}^{+q}) \over N({\rm X})}
    E_{ij} N_{\rm e} n_j A_{ji} \quad\quad {\rm erg~cm}^{-3}~{\rm s}^{-1}
\end{equation}
where $N$ denotes the number density of the relevant species and
$E_{ij}$ is the energy separation of the two atomic levels. This
equation is often rewritten as:
\begin{equation}
\epsilon_{ij}(N_{\rm e},T) = G(N_{\rm e},T) N_{\rm e}^2 
\end{equation}
where $G$ is the \emph{contribution function}.
Tables of the
element abundance relative to hydrogen, $N(X)/N_{\rm H}$, from various
sources are provided in CHIANTI. The ionization fractions,
$N({\rm  X}^{+q})/ N({\rm X})$, often written as $F(T)$, are computed
from ionization and recombination rates stored in CHIANTI and a table
of values 
for a wide range of temperatures is distributed in the database (see
Sect.~\ref{sect.ionbal} for more details). The hydrogen to electron ratio,
$N_{\rm H}/N_{\rm e}$, is calculated in a self-consistent manner using
the abundance and ion fraction tables in CHIANTI.  From
the emissivities, synthetic spectra can be computed by assuming a
model of the density and temperature structure of the plasma. See
Phillips et al.~\cite{phillips08} and the CHIANTI User Guide\footnote{Available at
  http://chiantidatabase.org/chianti\_user\_guide.html.} for more details.

In addition to the atomic data, CHIANTI contains a comprehensive
software package for computing the quantities described above.
For the original database the software was written in the Interactive
Data Language (IDL),
and this continues to be maintained to the present. In 2010 a
version of the software written in the \emph{Python} language was introduced
and called \emph{ChiantiPy} and this is maintained alongside the IDL
version. Brief descriptions of each software package are given below.

\subsection{IDL software}

The CHIANTI IDL software routine \verb|pop_solver| solves the matrix
equations describing the level balance processes yielding the $n_j$
values discussed above. The contribution function is computed using
the routine \verb|gofnt|. Synthetic spectra are created with
the graphic user interface routine \verb|ch_ss|, which calls the lower
level routines
\verb|ch_synthetic| and \verb|make_chianti_spec|. Line
ratio diagnostics are a crucial component of UV spectroscopy and these
can be studied with the routines \verb|dens_plotter| and
\verb|temp_plotter| for density and temperature diagnostics,
respectively. For multi-thermal plasmas a differential emission
measure describes the distribution of plasma with temperature, and the
CHIANTI routine \verb|chianti_dem| enables this to be
computed. Further information about the IDL software is available
in the CHIANTI user guide.

\subsection{Python software}

The \emph{ChiantiPy} software package is written in \emph{Python}
(https://www.python.org), a free, modern, object-oriented computer
language.  \emph{ChiantiPy} provides the ability to use the CHIANTI database
to calculate the emission properties of coronal plasmas.  It does this
largely through an object-oriented approach.  The most basic
class/object is the ``ion'', and an example of how to create the object
for \ion{C}{iv} is 
\begin{verbatim}
> import chianti.core as ch
> myC_4ion = ch.ion('c_4',temperature, density)
\end{verbatim}
where the user has already specified the temperature and electron
density.  The ``ion'' object has methods such as populate, popPlot, intensity,
intensityRatio, freeFree, freeBound, and spectrum, among others.  One
can calculate the level populations of the ion and plot them: 
\begin{verbatim}
> myC_4ion.populate()
> myC_4ion.popPlot() 
\end{verbatim}
The values of the population are then stored in a dictionary
\verb|myC_4ion.Population| which is directly available to the user.  Until
the user deletes the object, it and all of its methods remain for
further calculations. 

There are three spectrum classes: ``spectrum'', ``mspectrum'', and
``ipymspectrum''.  These all calculate the emission spectrum as a
function of temperature and density for a given wavelength and line
shape filter.  This is done through a collection of ``ion'' objects.
``spectrum'' is a single processor version and ``mspectrum'' and
``ipymspectrum'' are multiprocessor versions.  ``ipymspectrum'' can run
inside a \emph{Jupyter}/\emph{IPython} (http://ipython.org/) notebook or qtconsole.
``mspectrum'' can run in a \emph{Python} script or in a simple \emph{Jupyter}/\emph{IPython}
console. 

Further documentation on \emph{ChiantiPy} can be found at
http://chiantipy.sourceforge.net and the software package can be
downloaded from https://sourceforge.net/projects/chiantipy.  In the
near future, \emph{ChiantiPy} will be moving to \emph{GitHub} (https://github.com).

\section{Database structure}

The CHIANTI atomic data files are all plain ascii files lying under a top
directory called \verb|dbase|. The data files for ions are stored in
directories such as \verb|dbase/o/o_6| for \ion{O}{vi}. Each file
has a  name of the form \verb|o_6.[ext]| and the complete list of possible file
extensions, \verb|[ext]|, are given in Table~\ref{tbl.file}.

Additional data-sets distributed with CHIANTI include: tables of
commonly-used element abundance data-sets; a table of equilibrium
ionization fractions; and data-sets for computing continuum
emission. These are stored in the sub-directories \verb|abundance|,
\verb|ioneq| and \verb|continuum| of \verb|dbase|.

The following sections discuss the different types of atomic data-set
contained in CHIANTI.

\begin{table}[h]
\caption{CHIANTI file types.}
\begin{center}
\begin{tabular}{llp{11.5cm}}
\noalign{\hrule}
\noalign{\smallskip}
\noalign{\hrule}
\noalign{\smallskip}
~ &Extension  & Purpose \\
\noalign{\hrule}
\noalign{\smallskip}
\multicolumn{3}{l}{\emph{Level balance files}} \\
&ELVLC  & Level identifications and energies \\
&SCUPS & Scaled effective collision strengths and dielectronic capture
rates (for ``d'' files) \\
&WGFA & Decay rates, 2-photon rates, autoionization rates \\
&FBLVL & Configuration energies for free-bound continuum calculation \\
&PSPLUPS  & Spline fits to proton rate coefficients \\
&RECLVL & Level-resolved radiative recombination rates \\
&CILVL & Level-resolved ionization rates\\
\noalign{\smallskip}
\multicolumn{3}{l}{\emph{Ion balance files}} \\
&DIPARAMS & Spline fits to direct ionization cross-sections\\
&EASPLOM & Spline fits to excitation-autoionization cross-sections\\
&EASPLUPS & Spline fits to excitation-autoionization rates\\
&RRPARAMS & Fit parameters for radiative recombination rates\\
&DRPARAMS & Fit parameters for dielectronic recombination rates\\
\noalign{\hrule}
\end{tabular}
\end{center}
\label{tbl.file}
\end{table}

\subsection{Energy levels}

The ELVLC files contain a list of fine structure levels
for each ion, giving transition information and the observed and/or
theoretical energy. Each level is assigned an integer index, with the ground
level assigned 1, and these indices are used to identify the levels in
other data files. A level is only included in the file if there exist
excitation and decay data to enable the level's population to be
modeled.

There are two energy columns in the files: one for experimental
energies and the other for theoretical energies. Both are given in
units of cm$^{-1}$. 
The NIST database is the principal source of experimental
energy values, although other values are used where necessary. There are
often energy levels for which no experimental values are available and
so for these theoretical values are used from published
calculations. In some cases the theoretical values can be improved,
e.g., if another level in the same multiplet has a known energy, and
the CHIANTI team will compute a ``best-guess'' energy and insert this
into the theoretical energy column.

The format of the ELVLC files was changed in CHIANTI 8
\cite{chianti8}, giving a simpler structure to the files but
retaining the same information.

\subsection{Electron excitation rates}\label{sect.exc}

The electron excitation rates form the single most significant
data-set in CHIANTI and it is
the one to which the most effort is applied. The data are stored as
effective collision strengths, $\Upsilon$, (often referred to
simply as ``upsilons'') a dimensionless
number derived from the integral of the collision cross-section with the
Maxwellian distribution. Sect.~4.2.1 of Phillips et al.\ \cite{phillips08} gives
further details, including the expression that relates the $\Upsilon$
to the electron excitation rate coefficient, $C$, discussed earlier.

The effective collision strengths are not stored directly but instead
a scaling is applied, converting $\Upsilon(T)$ to $\Upsilon_{\rm
  s}(T_{\rm s})$, where $T_{\rm s}$ is a scaled temperature taking
values between 0 and 1, with 1 corresponding to $T=\infty$.  The
scaling formulae vary according to the type of the transition,
and the types are identified in Table~\ref{tbl.type}. Types 1--4
were introduced by Burgess \& Tully \cite{burgess92}, Type~5 was introduced in CHIANTI
3 \cite{chianti3}, and Type~6 in CHIANTI 4 \cite{chianti4}.

\begin{table}[h]
\caption{SCUPS file transition types.}
\begin{center}
\begin{tabular}{lp{8cm}}
\noalign{\hrule}
\noalign{\smallskip}
\noalign{\hrule}
\noalign{\smallskip}
Index  & Type \\
\noalign{\hrule}
\noalign{\smallskip}
1 & Allowed transitions \\
2 & Forbidden transitions \\
3 & Intercombination transitions \\
4 & Allowed transitions with a small $gf$ value \\
5 & Dielectronic recombination transitions\\
6 & Proton rates \\
\noalign{\hrule}
\end{tabular}
\end{center}
\label{tbl.type}
\end{table}

The scaled temperatures and upsilons are stored in the SCUPS files, and
they include the values $\Upsilon_{\rm  s}(0)$ and $\Upsilon_{\rm
  s}(1)$. The former are obtained by performing a backwards 
extrapolation in $T_{\rm s},\Upsilon_{\rm s}$ space. The infinite
temperature point for dipole-allowed transitions has a well-defined value
based on the transition energy and $gf$ value \cite{burgess92}, and
this is used for most ions. The high temperature limit for 
non-dipole allowed transitions can be calculated using the method of
Burgess et al.\ \cite{burgess97} (see also Whiteford et
al.\ \cite{whiteford01}), but most electron collision 
calculations do not provide these data. Exceptions are the recent
calculations performed for the Atomic Processes for Astrophysical
Plasmas (APAP) network, e.g., Liang et al.\ \cite{2009A&A...499..943L}.

Prior to CHIANTI 8 \cite{chianti8}, 5-point or 9-point splines were
fit to the scaled upsilon data but this method was abandoned as 
the splines could not reproduce the complete set of upsilons for some
transitions, and so it was necessary to remove some data points in
order to obtain a good fit. Usually the removed points were in the low
temperature regime, and so did not affect rates for electron-ionized plasmas
but they could be significant for photoionized plasmas. This change
has resulted in the scaled upsilons being stored in new SCUPS files,
replacing the older SPLUPS files.

Most of the electron excitation rates are for levels below the
ionization threshold of the ion, but for many ions we also include a
set of levels above the ionization threshold, corresponding to inner
shell excitation. For example, for the lithium-like ions with ground
configuration $1s^22s$, we include levels of the form $1s2s2l$
($l=s,p$) corresponding to an excitation of a $1s$ orbital. These
excitations are important in generating X-ray satellite lines (see
also Sect.~\ref{sect.diel}).

For most modern data-sets atomic physicists provide collision
cross-sections in the form of effective collision strengths that are
directly input to CHIANTI after scaling. In some cases the collision
strengths, $\Omega$ (often referred to as ``omegas''), are provided as
a function of energy, and these are 
integrated over a Maxwellian distribution by the CHIANTI team to yield
the upsilons that are then input to CHIANTI.

\subsection{Radiative decay rates, two-photon rates and autoionization
rates}\label{sect.wgfa}

The second key data-set for CHIANTI are the spontaneous radiative
decay rates (often referred to as ``$A$-values''),
which are essential for solving the level balance
equations. The rates are stored in the CHIANTI WGFA files together
with the weighted oscillator strengths ($gf$) and the wavelengths for
the transitions. The wavelengths are computed using the energies from
the ELVLC file. If either or both of the two levels involved in a
transition have theoretical energies, then the wavelength is given
as a negative number, which serves as a flag to the software to
indicate that the wavelength may not be accurate.

In addition to the spontaneous radiative decay rates, the WGFA files
are also used to store two other types of rate: two-photon decay rates
for hydrogen and helium-like ions, and autoionization rates.  The
two-photon decays are discussed in Sect.~\ref{sect.two-photon} and
enable two states in the hydrogen and helium-like ions to decay that
otherwise would be strictly forbidden, or decay very weakly. They
result in a continuum of emission discussed in
Sect.~\ref{sect.two-photon}, and they also need to be included in the
level balance of the ions. The critical difference in terms of the
entries in the WGFA file is that the two-photon decays are assigned a
zero wavelength so that a spectral line emissivity does not arise
from the transition, instead the emission is separately modeled via
the two-photon continuum emissivity calculation (Sect.~\ref{sect.two-photon}).

For many ions in CHIANTI we include atomic levels that lie above the
ionization threshold of the ion as these are needed for modeling
dielectronic recombination lines (principally for X-ray spectra modeling). These
levels can decay by regular spontaneous radiative decay, but they can
also decay through autoionization, i.e., spontaneous ejection of the
outermost electron. In terms of the emissivity calculation of the
ion, the autoionizations serve to reduce the population decaying to
lower levels in the ion. As for the two-photon decays, an autoionization
is represented with a zero wavelength for the transition. The
autoionization rate is inserted in the same column as the
$A$-value and identified as a transition direct to the ground
state. The transition of an autoionizing level to the ground state may
thus be represented twice in the WGFA file: once for the radiative
decay, and again for the autoionization rate.
For solving the level balance equations the two rates are
summed, but only the radiative decay yields an emissivity value.

\subsection{Proton excitation rates}\label{sect.proton}

Atomic levels can be excited by collisions with protons, but only if
the energy separation of the levels is small, as first demonstrated
for the ground transition of \ion{Fe}{xiv}
\cite{1964MNRAS.127..191S}. Even for cases where the proton rates are
larger than the electron rates, the dominant excitation process will
usually be cascading from higher levels and so proton rates generally do not
have a critical importance for the level balance equations. Most
proton rate calculations were performed in the 1970's through to the
1990's, and rates for CHIANTI were assessed and added in version~4 of
the database \cite{chianti4}.

The proton rates are stored in PSPLUPS files, which contain 5 or
9-point spline fits to the rate coefficients. Where possible the
transitions were fit with one of the electron excitation fitting
formulae (types 1--4 in Table~\ref{tbl.type}), but for many it was
necessary to perform a fit to the logarithm of the rates corresponding
to a type~6 transition \cite{chianti4}. For any single ion the PSPLUPS
files contain at most a 
handful of transitions.

\subsection{Dielectronic capture}\label{sect.diel}

CHIANTI 3 \cite{chianti3} extended CHIANTI to the
X-ray wavelength range and a particular focus was on the addition of
satellite lines. As an example, consider the strong $1s^2$ $^1S_0$ -- $1s2p$
$^1P_1$ transition of helium-like ions, and the transition $1s^23d$ --
$1s2p(^1P)3d$ in lithium-like ions. The latter is essentially
the same transition as for the helium-like ion, only it takes place in
the presence of a high-lying $3d$ electron. The $3d$ electron serves
to make a perturbation to the wavefunctions of the two states,
meaning the wavelength of the
lithium-like transition will be close to the helium-like transition,
hence it is referred to as a \emph{satellite line} of the helium-like
transition.

Continuing this example, the $1s2p(^1P)3d$ level lies above the
ionization threshold and it is excited either by excitation of the
inner $1s$ shell  or by dielectronic capture of a free electron onto
the helium-like system. The former case is a regular electron
excitation and is included in the SCUPS file
(Sect.~\ref{sect.exc}). The latter is a 
different process and is treated in the following way.

A completely new set of ion models were introduced with CHIANTI 3
\cite{chianti3} that
were identified by adding a ``d'' to the ion names. For example,
``o\_6d'' for the dielectronic files of \ion{O}{vi}. For each
dielectronic ion model
there is a ELVLC, WGFA and SCUPS file. The excitations in the SCUPS
files are actually dielectronic excitations coming from the
recombining ion and are fit as Type~5 transitions
(Table~\ref{tbl.type} and \cite{chianti3}). They are considered as
excitations from the ground level of the recombined ion, and only
excitations to the doubly-excited states are included in the file. The
WGFA file contains radiative decay rates and autoionization rates for
de-populating the doubly-excited levels. Note that a full set of decay
rates is needed in order to track the cascading through the ion's
level structure.

The dielectronic model produces a ``second spectrum'' for the ion
containing both new lines and duplicates of lines in the main ion
model. When computing the complete synthetic spectrum from a plasma,
the CHIANTI software sums the two spectra, thus potentially enhancing
lines from the original model.

\subsection{Level-resolved ionization and radiative recombination rates}

The processes of ionization and radiative recombination can leave the
final ion in an excited state, thus providing an additional population
component to the excited levels that enhances the intensities of emission
lines from these levels. To model these processes correctly it would
be necessary to develop models for the entire sequence of ions of an
element and allow transitions between states in neighboring ions, rather
than just within a single ion. This would be a significant change to
the structure of CHIANTI, greatly increasing processing time for the
CHIANTI software.

A convenient solution can be made in the case where the population of
the ground state of an ion is much greater ($\sim$~a factor 100) than
the population of any other state, and this was introduced in CHIANTI
5 \cite{chianti5} for the iron ions \ion{Fe}{xvii--xxiii} and selected
ions of the hydrogen and helium-like sequences.

Rather than include the ionization and recombination rates when
computing the level balance equations, the rates are introduced as
part of  a
correction factor applied to populations computed using the regular
CHIANTI calculation. For the ions considered, the method is accurate
up to densities of $\approx$ $10^{13}$~cm$^{-3}$.

The rate coefficients for the processes  are stored in RECLVL and CILVL files for recombination and
ionization, respectively, as a function of $\log\,T$. They are
considered as ``excitations'' from the ground level.

We highlight that the recombination data stored in the RECLVL file are for
radiative recombinations into bound levels of the ion. Recombinations into
doubly-excited states of the ion are modeled through the dielectronic
models of the ions (Sect.~\ref{sect.diel}). 

\subsection{Photoexcitation and stimulated emission}

In the presence of a radiation field the additional radiative
processes of photoexcitation and stimulated emission can take place,
however these do not require additional atomic data: the rates are
computed using the radiative decay rates (Sect.~\ref{sect.wgfa}) and a function representing
the energy density distribution of the radiation field. CHIANTI 4
\cite{chianti4} introduced modifications to the CHIANTI software to
allow the processes to be modeled assuming a blackbody radiation
field. A further modification was introduced in CHIANTI 5
\cite{chianti5} to allow an arbitrary radiation field to be input.

\section{Continuum emission}

Continuum emission is important for high temperature astrophysical plasmas
and for many years the standard reference for astronomers was
the work of Mewe et al.\ \cite{mewe86}, which is implemented in the IDL routine
CONFLX. The three components to the continuum are free-free,
free-bound and two-photon, although the latter is a minor contributor.
The current CHIANTI implementations were introduced in CHIANTI 4
\cite{chianti4}, and comparisons with Mewe et al.\ \cite{mewe86} were presented
by Landi \cite{2007A&A...476..675L}.
Data files for the continuum processes are stored in \verb|dbase/continuum|.

\subsection{Free-free}

Free-free or \emph{bremsstrahlung} emission occurs when a free
electron is decelerated during a collision with a positively-charged
ion, and it is typically implemented in spectral codes through
tabulations of the free-free Gaunt factor. For CHIANTI we use the
relativistic Gaunt factor tabulation of Itoh et al.\ \cite{itoh00}, supplemented
by the non-relativistic Gaunt factors of Sutherland \cite{sutherland98} for
parameter ranges not covered by Itoh et al.~\cite{itoh00}.

\subsection{Free-bound}

The free-bound emission results from the capture of a free electron by
an atom or ion and the capture can take place into any bound level of the
ion. The critical atomic parameter is the capture cross-section into a
bound state $i$, although in practice the cross-section for the
inverse process, photoionization, is computed with the two
cross-sections related by the principle of detailed balance (the Milne
relation). The expression relating the photoionization cross-sections
to the free-bound emissivity is given in Eq.~12 of Young et al.~\cite{chianti4}.

For CHIANTI the photoionization cross-sections from the ground state
are taken from Verner et al.~\cite{verner95}, and for excited levels (i.e., excited
levels in the recombined ion) the Gaunt factor expression of Karzas \& Latter
\cite{karzas61} is used. For this expression levels are treated as
configurations, thus if the ground configuration is $3s^23p^2$, then
the excited configurations are $3s^23pnl$ and Gaunt factors are
available for $n$ up to 6 and $l$ up to 5. Energies for these
configurations are stored in the CHIANTI .FBLVL files, which are
available for each ion. See Young et al.~\cite{chianti4} for more details.

The data files for the Verner et al.\ and Karzas \& Latter data-sets
are stored in \verb|dbase/continuum| and IDL routines
(\verb|verner_xs| and \verb|karzas_xs|) are available
for deriving the photoionization cross-sections from these two
data-sets.

\subsection{Two-photon}\label{sect.two-photon}

The two-photon continuum arises from the decays of the $1s$
$^2S_{1/2}$ state in hydrogen-like ions, and the $1s2s$ $^1S_0$ state
in helium-like ions. The transition from the helium-like state to the
ground is strictly forbidden and
so the two-photon decay is the only decay route, whereas the
hydrogen-like state has a weak magnetic dipole decay to the ground but
this is weaker than the two-photon decay rate for most ions.

The decay route in both cases is the simultaneous emission of two
photons for which the summed energies correspond to the energy
difference between the excited state and the ground. Other than this
restriction, the photons can take any energy and thus a continuum
results. The key atomic parameters are the decay rate and the spectral
distribution function, which describes the distribution of photon
energies for an ensemble of particles. These parameters are stored in
data-files in the \verb|dbase/continuum| directory of CHIANTI. More details
are given in Young et al.~\cite{chianti4}.

The decay rates are also stored in the WGFA files as they are necessary
for solving  the level balance equations for the hydrogen and
helium-like ions (see Sect.~\ref{sect.wgfa}).

\section{Ionization balance}\label{sect.ionbal}

The major additions to CHIANTI 6 were total  ionization and
recombination rates for all ions, allowing the equilibrium ionization balance
of the plasma to be computed for any temperature. These equations
simply need total rates between neighbouring ions along an element's
ionization sequence, and they yield the ionization fractions, $F(T)$,
for each ion. Since ionization and recombination are electron
collision processes then, to first order, there is no density
dependence to the ionization fractions. However, as density increases
then metastable level populations can become significant and give different
routes for ionization and recombination to take place, modifying the
total ionization and recombination rate coefficients. In addition,
dielectronic recombination is known to be suppressed at high densities
(e.g., Nikoli\'c et al.~\cite{nikolic13}).
For CHIANTI we assume the zero density approximation, and use total,
density-independent ionization and recombination rates for determining
the ion balance.
Prior to the CHIANTI 6 update
astrophysicists relied on occasional updates to  the zero-density ionization
balance calculations, the most well-known being those of Arnaud \& Rothenflug
\cite{1985A&AS...60..425A}, Arnaud \& Raymond
\cite{1992ApJ...398..394A} and Mazzotta et al.\ \cite{mazzotta98}. The
addition to CHIANTI allows the ionization fractions to be updated on
the typical $\approx$~2 year update schedule of the database, and
researchers are recommended to use the CHIANTI calculations as these
contain improved atomic data compared to the earlier calculations.

The formats for storing the recombination and ionization rates are
described below.

\subsection{Ionization rates}

Ionization includes both direct ionization and
excitation-autoionization.  The parameters for calculating the direct
ionization cross-sections are contained in the DIPARAMS file. Dere~\cite{2007A&A...466..771D}
developed a Burgess \& Tully  \cite{burgess92} type of scaling for direct
ionization cross-sections.  The data in these files are fits to the
scaled energies and ionization cross-sections.  These may be either
experimental cross-sections or calculated cross-sections.  The direct
ionization from several shells can be important and this is reflected
by fits to more than a single cross-section. For example,
\ion{Fe}{xiii} has a ground configuration of $2p^63s^23p^2$ and the
DIPARAMS file
contains fits to the three sets of cross-sections for ionizations of the $3p$, $3s$
and $2p$ orbitals. The total cross-section is obtained by summing
these cross-sections.  The direct ionization
rate coefficients are obtained by a 12 point Gauss-Laguerre
integration over a Maxwellian electron distribution. Full details are
given by 
Dere~\cite{2007A&A...466..771D}.

Excitation-autoionization occurs when an electron collides with an ion
and excites it into a state above the ionization potential.  The ion in
this state can undergo a stabilizing radiative transition, leaving it
in a stable state of the original ion.  If the excited ion undergoes
autoionization then it is left in a stable state of the higher
ionization stage.  The parameters for calculating the ionization
cross-section by excitation-autoionization are in the EASPLOM
files.  These contain Burgess-Tully fits to the excitation
cross-sections multiplied by the branching ratio of the autoionization
process appropriate for the excited state.  The parameters for
calculating the excitation-autoionization rate coefficients are
developed by an integration over a Maxwellian velocity distribution.
The results are stored as Burgess-Tully fits to the scaled rate
coefficients.  The parameters for calculating the ionization rate
coefficients by excitation-autoionization are in the EASPLUPS
files. Considering ionization of \ion{Fe}{xiii} again, the excitation
routes considered by Dere~\cite{2007A&A...466..771D} are $n=2$ to $n=3$, 4
and 5,  and so  there are three entries in the
EASPLOMS and EASPLUPS files.

\subsection{Recombination rates}

Recombination is the capture of an electron into a
bound state of the recombined ion. The capture can take place either
directly, or indirectly by
capture to an unstable doubly-excited state of the recombined ion,
followed by a radiative decay to a bound state. The two types are
referred to as radiative and dielectronic recombination, respectively,
and in terms of theoretical calculations they are usually computed
separately. One exception is the method of Nahar \& Pradhan~\cite{1992PhRvL..68.1488N}
which computes the total recombination coefficient in a single
calculation, however Pindzola et al.~\cite{1992PhRvA..46.5725P} have demonstrated
that 
interference between the RR and DR processes is negligible, justifying
the independent processes approach.

\subsubsection{Radiative recombination}

Radiative recombination (RR) data are stored in the RRPARAMS files, which
contain a single set of fit parameters for the total RR rate coefficient.
There are three types of fitting formulae used for RR rates which were
introduced by Aldrovandi \& Pequignot \cite{1973A&A....25..137A},
Verner \& Ferland \cite{1996ApJS..103..467V}, and Gu
\cite{2003ApJ...589.1085G}.

The bulk of the RR data used in CHIANTI are from recent calculations of N.R.~Badnell
and collaborators. Badnell \cite{badnell06} performed calculations for all elements up to
zinc for all ions from bare nucleus to sodium-like. Further work for
the magnesium, aluminium and argon sequences were performed by Altun
et al.\ \cite{2007A&A...474.1051A}, Abdel-Nagy et
al.\ \cite{2012A&A...537A..40A} and Nikoli\'c et
al.\ \cite{2010A&A...516A..97N}, and data for additional iron ions
were presented by Badnell \cite{2006ApJ...651L..73B} and
Schmidt et al.\ \cite{2008A&A...492..265S}. Data for the remaining
ions are mostly from older calculations and are summarized in Dere et
al.~\cite{chianti6}.

\subsubsection{Dielectronic recombination}

Dielectronic recombination (DR) was described previously 
(Sect.~\ref{sect.diel}) in regard to modeling the strength of satellite
lines at X-ray wavelengths. For modeling the ionization balance the
total DR rate coefficients are required and a recent project described
by Badnell et al.~\cite{2003A&A...406.1151B} produced complete sets of
rates for all ions up to zinc for all sequences from hydrogen-like to
aluminium-like. Beyond these sequences there are few calculations in
the literature and rates are mostly calculated with  variants of the Burgess
General Formula \cite{1965ApJ...141.1588B} or interpolation -- see
Dere et al.~\cite{chianti6} for more details.

All DR rates are fit with a standard formula first given by Arnaud \&
Raymond~\cite{1992ApJ...398..394A}, and the CHIANTI DRPARAMS file contains
the fit parameters.

\section{Applications}

The previous sections described the contents of the CHIANTI database,
and in this section we present five examples of the varied uses of
CHIANTI within the solar physics community.

\subsection{Density diagnostics}

For many ions one can find pairs of emission lines that are sensitive
to the electron density, a famous example being the \ion{O}{ii}
\lam3729/\lam3726 ratio (Seaton \& Osterbrock
\cite{seaton57}). Density diagnostics are among the most important
applications of 
CHIANTI, and two examples are illustrated here from the \emph{Hinode}/EIS
and Interface Region Imaging Spectrometer (IRIS; \cite{iris})  missions.

\begin{figure}[h]
\centerline{\epsfxsize=6in\epsfbox{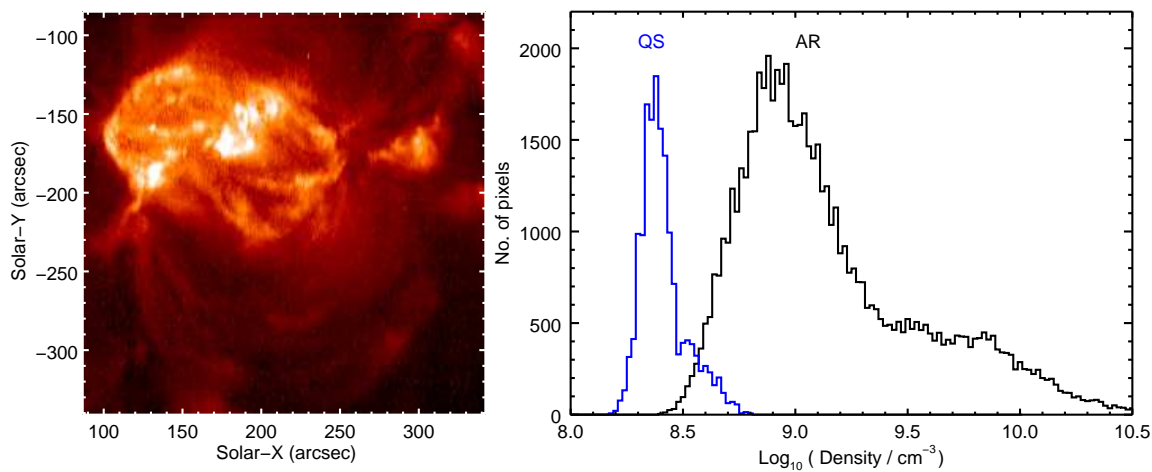}}
\caption{The left panel shows a density map with logarithmic scaling
  derived from an EIS raster beginning at 10:33~UT on 2006 December
  2. The map is derived from \ion{Fe}{xii} \lam186.9/\lam195.1. The
  right panel shows a histogram of the derived densities from the
  active region (AR). The blue histogram shows densities derived from
  the same ratio from a quiet Sun (QS) raster beginning at 12:15~UT on 2010
October~8.}
\label{fig.eis-dens}
\end{figure}

One of the most important density diagnostics for the
\emph{Hinode}/EIS mission is \ion{Fe}{xii} \lam186.9/\lam195.1
(\lam186.9 is actually a blend of two close transitions) that is
formed at 1.5~MK. The ratio yields very precise density measurements
(Young et al.~\cite{young09-dens}) enabling high quality density maps of
active regions, such as shown in
Figure~\ref{fig.eis-dens}. Comparisons of densities in this active
region with a typical quiet Sun data-set show that coronal loops in
the periphery of  the
active region have a density $\approx$0.5~dex larger than quiet Sun,
whereas bright ``knots'' of emission in the active region core can be
1.5--2.0 dex higher. Atomic data from CHIANTI 8 were used to derive
the densities and the EIS calibration of Del Zanna~\cite{dz-cal} was used.

\begin{figure}[h]
\centerline{\epsfxsize=10cm\epsfbox{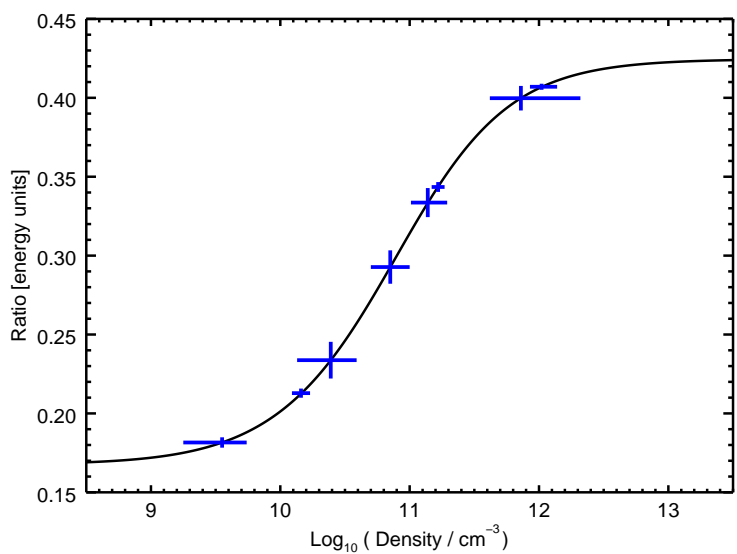}}
\caption{The solid line shows the theoretical variation of the
  \ion{O}{iv} \lam1399.8/\lam1401.2 ratio as a function of density
  computed at $\log\,T=5.15$ using CHIANTI 8. The blue crosses show
  the measured ratios and densities with 2-$\sigma$ error
  bars. Figure adapted from Young~\cite{young15-o4}.}
\label{fig.o4-dens}
\end{figure}

\ion{O}{iv} yields a set of five intercombination transitions between
1397 and 1407~\AA\ that are widely used in both solar physics and
astrophysics (e.g., Hayes \& Shine~\cite{hayes87}; Keenan et al.~\cite{keenan02}). IRIS is the most
recent UV instrument to measure these lines, and 
Figure~\ref{fig.o4-dens} shows the theoretical variation of the
\ion{O}{iv} \lam1399.8/\lam1401.2 ratio obtained from CHIANTI 8 with observed ratio values from IRIS
over-plotted. 
The measurements were obtained from a range of solar features
(Young~\cite{young15-o4}), and the largest values are obtained from bright
flare kernels, while the lowest value is obtained from a coronal
hole. This plot demonstrates both that the diagnostic is very useful
in a wide range of conditions, but also that the atomic data in
CHIANTI allows accurate measurements of the density.

\subsection{Reponse functions for the SDO/AIA instrument}

\begin{figure}[h]
\centerline{\epsfxsize=6in\epsfbox{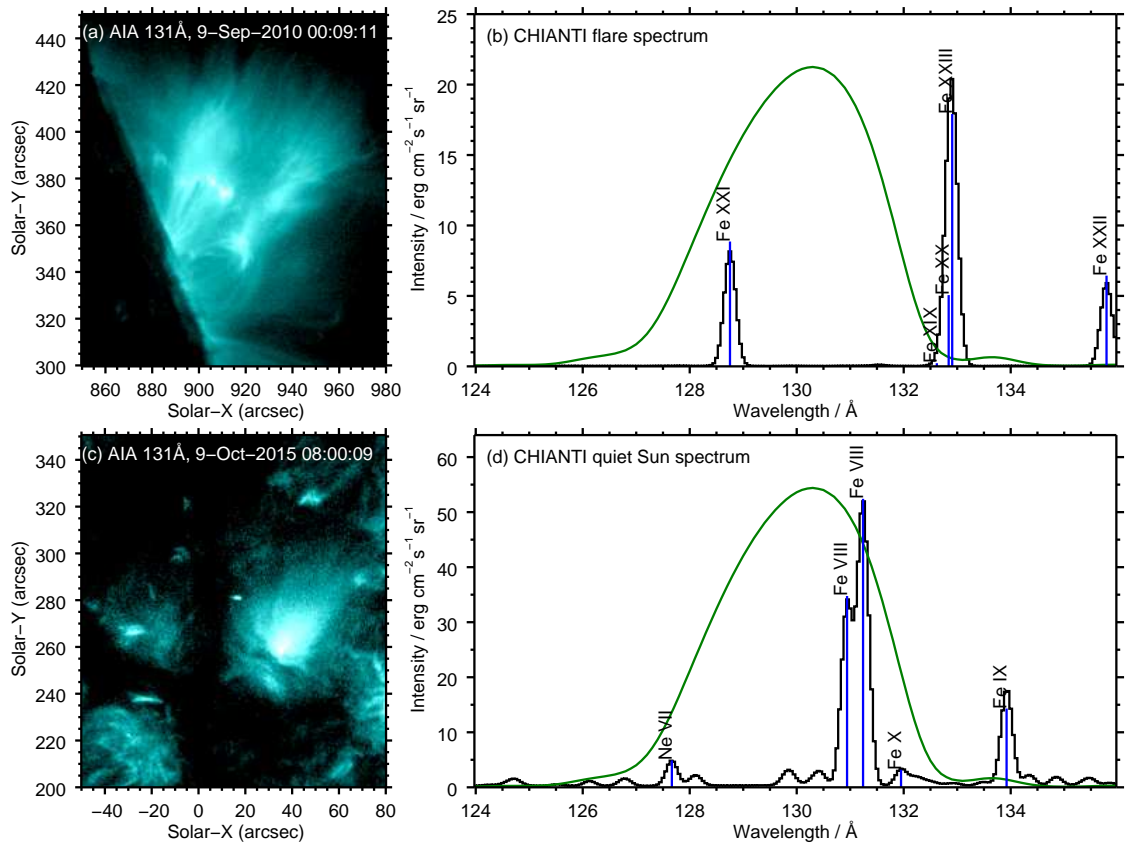}}
\caption{Panels a and c show images obtained with the AIA
  131~\AA\ filter, both displayed with a logarithmic intensity
  scaling. Panel a shows a flare at the solar limb, and panel c shows
  a bright plume within a low latitude coronal hole. Panels b and d
  show synthetic CHIANTI spectra for flare and quiet Sun conditions,
  respectively, with the AIA 131~\AA\ response function over-plotted
  in green. The five strongest lines in each spectrum are indicated.}
\label{fig.a131}
\end{figure}

The AIA instrument on board the SDO satellite has seven EUV filters
that pick out different wavelength regions in the solar spectrum,
giving the instrument a wide temperature coverage (O'Dwyer et
al.~\cite{odwyer10}). 
One of the filters is centered at 131~\AA\ and CHIANTI is used by
the AIA team to model the response of the filter to different solar
conditions, as illustrated in Fig.~\ref{fig.a131}. Panel a
shows an image recorded from a solar flare that occurred at the solar
limb. In these conditions the underlying solar spectrum, as modeled
with CHIANTI, is shown in panel b. Although the strongest transition
is \ion{Fe}{xxiii} \lam132.91, the instrument response function (shown
in green) is low at this wavelength and the dominant transition is
actually \ion{Fe}{xxi} \lam128.75, formed at 11~MK.

Fig.~\ref{fig.a131}c shows an image obtained in a coronal hole,
showing a number of plume structures. In these conditions the CHIANTI
spectrum is as shown in Fig.~\ref{fig.a131}d, with all of the hot
flare lines absent and instead the dominant transitions come from
\ion{Fe}{viii}, formed at 0.8~MK.

\subsection{Modeling of the solar corona}\label{sect.model}

\begin{figure}[h]
\centerline{\epsfxsize=12cm\epsfbox{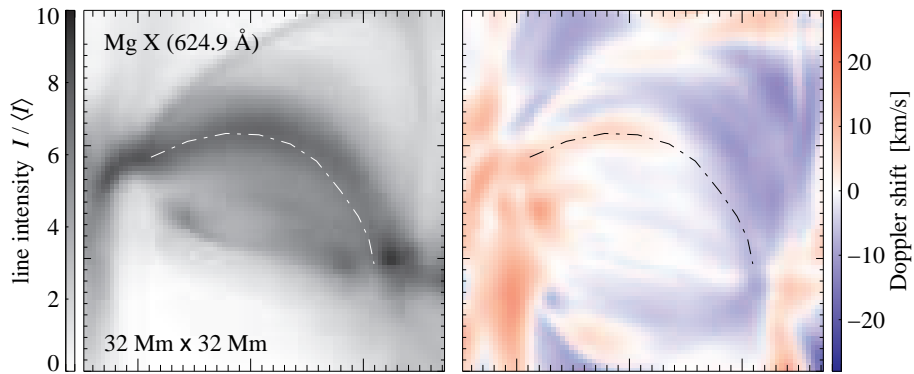}}
\caption{Coronal emission synthesised from a snapshot of a 3D MHD
  model of an active region. Courtesy of Hardi Peter.}
\label{fig.hardi}
\end{figure}

Modern computing power enables 3D magnetohydrodynamic (MHD) models of
the solar corona to be constructed, and often CHIANTI is used to
synthesize observed emissions from the models. An example is shown in
Figure~\ref{fig.hardi}, which is taken from Peter~\cite{peter10} and
uses the model presented by Peter et al.~\cite{peter06}. It shows a
view from the top of the computational box which corresponds to an observation near
the center of the solar disk. 
The left panel shows the line intensity as expected for \ion{Mg}{x} (624.9~\AA) originating from plasma at about 1 MK. Synthesising spectral line
profiles at each grid point of the computational domain, line-of-sight
integration (here along the vertical) provides a map of line profiles
similar to what is acquired during a raster scan with a slit
spectrograph. Fitting a single Gaussian to the line profiles then
provides a Doppler map (right panel). At this particular instance in
time a bright loop can be seen that hosts a siphon flow from the right
to the left that shows up as a blueshift at the right and a redshift
at the left leg of the loop (indicated by the dashed lines). The
coronal emission was synthesised using an earlier version of CHIANTI
(v4.02; Young et al.~\cite{chianti4}). 




\subsection{Solar wind diagnostics}

The solar wind plays a critical role in shaping the heliosphere and in 
determining the motion, arrival time and geo-effectiveness of Earth-directed 
coronal mass ejections (CMEs), and thus is a fundamental ingredient in Space Weather predictions.
One of the main tools to investigate the properties, heating, acceleration
and origin of the solar wind is the charge state composition measured in 
situ by space instrumentation. This quantity is determined by two key 
factors: 1) the evolution of the wind plasma velocity,  electron density 
and temperature, and 2) the ionization and recombination rates of wind ions. 

CHIANTI provides all the ionization and recombination data necessary to
calculate the evolution of the ionization status of the plasma from the
wind source region to the freeze-in point. This quantity can be used for 
several purposes. For example, Schwadron et al.~\cite{2011ApJ...739....9S} used the O$^{7+}$/O$^{6+}$ 
ratio to provide a crude estimate of the temperature of the solar corona, 
suggesting that the ratio decrease during solar cycle 23 indicated a 
global cooling of the solar corona. Edgar \&
Esser~\cite{2000ApJ...532L..71E} and Ko et al.~\cite{1996GeoRL..23.2785K}  compared predicted and observed charge state composition to study 
nonthermal electrons in the solar atmosphere; Esser \&
Edgar~\cite{2001ApJ...563.1055E}  used 
charge states to investigate differential velocities in the wind; and 
Ko et al.~\cite{1997SoPh..171..345K} and Esser et al.~\cite{1998ApJ...498..448E} used the measured oxygen charge state 
distribution to provide empirical models of the evolution of the wind 
velocity, electron density and temperature with distance from the source 
region. 

CHIANTI has enabled a more accurate and comprehensive study of solar
wind ionization by providing a compact database with the necessary
rates. For example, Landi et al.~\cite{2012ApJ...761...48L}  used CHIANTI-based calculation
of the wind ionization status to systematically study departures from 
equilibrium in the wind plasma, finding that they are significant and 
affect spectral line emission even in the inner
corona and so need to be taken into account. Landi et
al.~\cite{2012ApJ...758L..21L} showed that 
the radiative losses of the wind depart from equilibrium at transition
region temperatures, altering the energy equation in solar wind models.
Landi et al.~\cite{2012ApJ...744..100L} also devised a new diagnostic technique which is based on
CHIANTI calculations of the wind plasma ionization to produce empirical
models of the wind velocity, electron density and temperature; in this 
technique in-situ measurements of the wind charge state composition
and remote sensing spectral observations of the wind source regions 
are compared with CHIANTI-based spectra and charge state distributions
calculated under non-equilibrium ionization.

\subsection{Applications to non-Maxwellian plasmas}

The assumption of a Maxwellian distribution for the particle energies
is fundamental to CHIANTI, encapsulated in the use of effective
collision strengths and rate coefficients for the electron and proton
collision processes (Sects.~\ref{sect.exc}, \ref{sect.proton} and \ref{sect.ionbal}). Extension to non-Maxwellians is possible if the
distribution can be expressed as a linear combination of Maxwellians
at different temperatures, and modifications to the CHIANTI software were
performed in CHIANTI 4 \cite{chianti4} to enable this. An example
of the use of the method was presented by Muglach et
al.~\cite{muglach10} who investigated how the effect of a
high-temperature tail of electrons would modify emission line ratios.

An alternative approach was presented by Dzif{\v c}{\'a}kov{\'a}~\cite{2006SoPh..234..243D} who provided a
method for deriving  collision strengths integrated over a kappa
distribution directly from the CHIANTI upsilon data. The validity of
this method was demonstrated by Dzif{\v c}{\'a}kov{\'a} \&
Mason~\cite{2008SoPh..247..301D} and it was  extended to
$n$-distributions by Dzif{\v
  c}{\'a}kov{\'a}~\cite{2006ESASP.617E..89D}. A database called KAPPA
\cite{2015ApJS..217...14D} was released in 2015 that effectively
provides an alternative version of CHIANTI for modeling spectra from
plasmas with kappa distributions. The atomic data are all derived from
the atomic data within CHIANTI. An application of these non-Maxwellian
data was presented by Dud\'ik et al.~\cite{2015ApJ...807..123D} who
were able to demonstrate that a transient coronal loop spectrum
obtained with the EIS instrument was consistent with a kappa electron
distribution with $\kappa\le 2$. 

\section{Summary}

The CHIANTI atomic database is very widely used in solar physics and
astrophysics, and the current status and contents have been
summarized. In addition examples of the diverse range of applications
have been presented. The CHIANTI team continues to respond to the needs
of the science community, and future updates will include
density-dependent ionization balance calculations and support for
non-equilibrium plasma studies. 

\ack

PRY, EL and KPD acknowledge support from NASA grant NNX15AF25G. GDZ
and HEM acknowledge support by the UK STFC 
and SOLID (First European SOLar Irradiance Data Exploitation),
a collaborative SPACE Project under the Seventh Framework Programme
(FP7/2007-2013) of the European Commission
under Grant Agreement No.~313188. Dr Hardi Peter is thanked for providing
Figure~\ref{fig.hardi} and text for Sect.~\ref{sect.model}.  This
research has made use of NASA's Astrophysics Data System Bibliographic
Services. 

\section*{References}

\bibliographystyle{iopart-num}
\bibliography{myrefs}

\end{document}